\documentclass[aps,prb,preprint,amsfonts,amssymb,a4paper,showkeys,showpacs,preprintnumbers]{revtex4-1}
\usepackage{ucs}
\usepackage[utf8x]{inputenc}
\usepackage[TS1,T1]{fontenc}
\usepackage{xspace}
\usepackage{hyperref}

\usepackage{amsmath}
\usepackage{amssymb}
\newcommand{\spmatrix}[1]{\left( \begin{smallmatrix}#1\end{smallmatrix}\right)} 
\newcommand\gC{{\mathbb C}}
\newcommand\gR{{\mathbb R}}
\newcommand\caA{{\mathcal A}}
\newcommand\caS{{\mathcal S}}
\newcommand\ksu{{\mathfrak{su}}}
\newcommand\bbbone{{ \mathchoice {1\mskip-4mu\mathrm{l} } {1\mskip-4mu\mathrm{l} }{1\mskip-4.5mu\mathrm{l} } {1\mskip-5mu\mathrm{l}} }}
\newcommand\Ad{{\text{\textup{Ad}}}} 

\newcommand{\au}{s} 
\newcommand{\asu}{u} 
\newcommand{\whS}{\widehat{s}}

\begin{document}

\title{A Remark on the Spontaneous Symmetry Breaking Mechanism\\ in the Standard Model}

\author{Thierry Masson}
\email{thierry.masson@cpt.univ-mrs.fr}
\affiliation{Centre de Physique Théorique (UMR 6207)\\
Case 907 - Campus de Luminy\\
F-13288 Marseille Cedex 9}

\author{Jean-Christophe Wallet}%
\email{jean-christophe.wallet@th.u-psud.fr}
\affiliation{Laboratoire de Physique Théorique (UMR 8627)\\
Bât 210, Université Paris-Sud Orsay\\
F-91405 Orsay Cedex}%

\date{January 8, 2010}

\begin{abstract}
In this paper we consider the Spontaneous Symmetry Breaking Mechanism (SSBM) in the Standard Model of particles in the unitary gauge. We show that the computation usually presented of this mechanism can be conveniently performed in a slightly different manner. As an outcome, the computation we present can change the interpretation of the SSBM in the Standard Model, in that it \emph{decouples} the $SU(2)$-gauge symmetry in the final Lagrangian instead of breaking it.
\end{abstract}

\preprint{CPT-P002-2010}
\preprint{LPT-Orsay 10-03}

\maketitle

The Spontaneous Symmetry Breaking Mechanism (SSBM) is an important ingredient of the Standard Model of particle physics. Until now, it is the only convincing procedure which permits one to implement mass generation to vector bosons and fermions without getting in trouble with renormalizability. 

From a technical point of view, what is called the unitary gauge has been used to exhibit and understand the particle content of this model in that it permits to write the Lagrangian in terms of fields which have definite charges under the remaining $U(1)$-gauge group.

In this note, we propose to reconsider the SSBM of the Standard Model in the unitary gauge. The computation we present seems to be a better way to manage this unitary gauge, in that on one hand it simplifies the presentation and on the other hand it makes manifest a possible different interpretation of this mechanism. Because this computation does not need any reference to the quantized version of the theory, we will explicitly stay at the classical level of the theory.

The salient feature of this computation is that the $SU(2)$ symmetry is not broken in the usual way but it is \emph{factored out} in the final Lagrangian. In that respect, it helps to display the reason why a part of the symmetry disappears without making some arbitrary choice, for instance for the fundamental configuration of some fields or for the specific value of the vacuum of the theory. As a direct consequence, the SSBM of the Standard Model should be better interpreted as a \emph{decoupling} of a symmetry. This computation also exhibits, as expected, the residual $U(1)$ symmetry and it shows how this group acts on the fields.

There is no new technicalities in this computation. At the classical level, the Lagrangians in the symmetric phase and in the broken phase are the usual ones. In our computation which connects these two Lagrangians, the key point is to take the unitary gauge for what it is from a mathematical point of view, and then to perform the necessary technical steps as they present themselves. As we shall see, some confusion may have been widespread in the literature about a so-called gauge transformation used in the computation. This point is clarified in the text because it is at the heart of the possible two routes (the usual one and our) which lead to the Lagrangian in the broken phase.

For explanatory reasons, this paper focuses only on a specific part of the Lagrangian of the Standard Model of particles. But in fact it can be easily seen that this computation can be adapted to the complete Lagrangian of the Standard Model, and to certain classes of Lagrangians where a SSBM occurs. A systematic and more mathematical study of this procedure is beyond the scope of this paper and is under study.

\section{The computation}
\label{thecomputation}

Let us first fix the notations and ingredients which will be used in the following. We will restrict ourselves to  the electroweak part of the Standard Model coupled to leptons and to only one flavor \cite{Glas:61a, *Wein:67a, *Sala:68a}. Indeed, the computation can be easily generalized to the complete Standard Model without difficulty. For standard textbooks, see \textit{e.g.} \cite{Wein:05b, *PeskSchr:08a, *AitcHey:04a, *ChenLi:84a}.

Let us stress once again that in what follows, we will always stay at the classical level of the theory.

The structure group is the usual $U(1) \times SU(2)$. For a group $G$, we denote by $\underline{G}$ its gauge group, \textit{i.e.} the group of smooth functions on space-time with values in $G$. The scalar Higgs fields doublet is denoted by $\phi = \spmatrix{\phi_1 \\ \phi_2}$. Left fermions define a $SU(2)$-doublet $\psi_L$ of (left projected) Dirac spinors and the right fermion is a singlet $\psi_R$, a (right projected) Dirac spinor. The gauge fields are denoted by $a_\mu$ for the gauge group $\underline{U(1)}$ and $b_\mu = b^a_\mu \sigma_a$ for the gauge group $\underline{SU(2)}$. Here the $\sigma_a$'s ($a=1,2,3$) are the standard Pauli matrices. The gauge fields are hermitean: $a_\mu^\dagger = a_\mu$ and $b_\mu^\dagger = b_\mu$.

The actions of the gauge groups $\underline{U(1)}$ and $\underline{SU(2)}$ on these fields are summarized in the following relations, with $\au \in \underline{U(1)}$ and $\asu \in \underline{SU(2)}$. In the following, we will always denote by $\au$ and $\asu$ the $U(1)$ and $SU(2)$ gauge transformations.
\begin{align*}
\phi^\au &= \au^{-1} \phi
&
\phi^\asu &= \asu^{-1} \phi 
\\
\psi_L^\au &= \au \psi_L
&
\psi_L^\asu &= \asu^{-1} \psi_L
\\
\psi_R^\au &= \au^{2} \psi_R
&
\psi_R^\asu &= \psi_R
\\
b_\mu^\au &= b_\mu
&
b_\mu^\asu &= \asu^{-1} b_\mu \asu + \tfrac{2i}{g} \asu^{-1} \partial_\mu \asu
\\
a_\mu^\au &= a_\mu + \tfrac{2i}{g'} \au^{-1} \partial_\mu \au
&
a_\mu^\asu &= a_\mu\,.
\end{align*}
The corresponding covariant derivatives are given by:
\begin{align*}
D_\mu \phi &= (\partial_\mu - i \tfrac{g}{2} b_\mu - i \tfrac{g'}{2} a_\mu) \phi
\\
D^L_\mu \psi_L &= (\partial_\mu - i \tfrac{g}{2} b_\mu + i \tfrac{g'}{2} a_\mu) \psi_L
\\
D^R_\mu \psi_R &= (\partial_\mu + i g' a_\mu) \psi_R\,.
\end{align*}

The part of the Lagrangian of the Standard Model we will consider in the following is given by
\begin{multline}
\label{eq-lagrangianbeginning}
\mathcal{L} = (D_\mu \phi)^\dagger (D^\mu \phi) - \mu^2 \phi^\dagger \phi - \lambda (\phi^\dagger \phi)^2\\
+ \overline{\psi_L} i \gamma^\mu D_\mu \psi_L + \overline{\psi_R} i \gamma^\mu D_\mu \psi_R + f (\overline{\psi_L} \phi \psi_R + \overline{\psi_R} \phi^\dagger \psi_L)\\
 -\frac{1}{4} f_{\mu \nu} f^{\mu \nu} -\frac{1}{4} \sum_a g^a_{\mu \nu} g^{a\; \mu \nu},
\end{multline}
where $\mu$ and $\lambda$ are the usual parameters for the potential of the scalar fields $\phi$, $f$ is a Yukawa coupling constant, $f_{\mu \nu}$ is the field strength of $a_\mu$ and $g_{\mu \nu}$ is the field strength of $b_\mu$ written as $g_{\mu \nu} = g^a_{\mu \nu}\frac{\sigma_a}{2}$.

The idea behind the unitary gauge is to remark that any non zero vector $\phi = \spmatrix{\phi_1 \\ \phi_2} \in \gC^2$ can be uniquely written as
\begin{equation}
\label{eq-phiUeta}
\phi = U \eta \begin{pmatrix} 0 \\ 1 \end{pmatrix},
\end{equation}
with $U \in SU(2)$ and $\eta \in \gR^+$ given by
\begin{align*}
U &= 
\begin{pmatrix} \overline{\phi_2}/\eta\rule[-10pt]{0mm}{0mm} & \phi_1/\eta \\ 
- \overline{\phi_1}/\eta & \phi_2/\eta \end{pmatrix},
&
\eta &= \sqrt{|\phi_1|^2 + |\phi_2|^2}\,.
\end{align*}
Notice that $U$ is not well defined at points $x$ for which $\eta(x) = 0$, as it is the case for polar-like coordinates. We will comment more on this after the end of the computation.

Now, passing to functions on space-time, we keep the reference vector $\spmatrix{0 \\ 1}$ constant, so that $U$ and $\eta$ are new field variables in place of $\phi_1$ and $\phi_2$. By construction, $U$ is a function with values in $SU(2)$ (possibly not defined at $x$ such that $\phi(x)=0$) and $\eta$ is a smooth function with values in $\gR^+$. 

This parametrization of the scalar fields $\phi$ is the usual way to perform the SSBM in the unitary gauge. The usual next step is to ``gauge away'' the $U$ fields, and to write down the residual Lagrangian. 

Here is the point where we diverge form this standard procedure. We will not gauge away the $U$ fields as it is commonly done. Indeed, we will consider them as dynamical variables submitted to the actions of the gauge groups. Doing that, it is possible to manage them through a convenient change of variables in such a way that they ``disappear''.

Because $U$ and $\eta$ are the new field variables for the scalar fields, they inherit well-defined transformations properties under the actions of the gauge groups. With the same notations as before, defining the transformed fields by $\phi^\asu = U^\asu \eta^\asu \spmatrix{0 \\ 1}$, one gets (recall that the reference vector $\spmatrix{0 \\ 1}$ is chosen once and for all)
\begin{equation*}
\phi^\asu = \asu^{-1} U \eta \begin{pmatrix} 0 \\ 1 \end{pmatrix} = (\asu^{-1} U) \eta \begin{pmatrix} 0 \\ 1 \end{pmatrix},
\end{equation*}
from which we deduce that under the action of the gauge group $\underline{SU(2)}$ one has
\begin{align*}
U^\asu &= \asu^{-1} U,
&
\eta^\asu &= \eta\,.
\end{align*}

Defining in the same way $\phi^\au = U^\au \eta^\au \spmatrix{0 \\ 1}$ one gets
\begin{align*}
U^\au &= 
\begin{pmatrix} \au \overline{\phi_2}/\eta\rule[-10pt]{0mm}{0mm} & \au^{-1}\phi_1/\eta \\ 
- \au \overline{\phi_1}/\eta & \au^{-1}\phi_2/\eta \end{pmatrix},
&
\eta^\au &= \sqrt{|\phi_1|^2 + |\phi_2|^2}\,.
\end{align*}
$U^\au$ can be uniquely written as a product of $U$ with an element of $\underline{SU(2)}$:
\begin{equation*}
U^\au = 
\begin{pmatrix} \overline{\phi_2}/\eta\rule[-10pt]{0mm}{0mm} & \phi_1/\eta \\ 
- \overline{\phi_1}/\eta & \phi_2/\eta \end{pmatrix}
\begin{pmatrix} \au \rule[-10pt]{0mm}{0mm} & 0 \\ 
0 & \au^{-1} \end{pmatrix}
=
U \whS,
\end{equation*}
with
\begin{equation*}
\whS = \begin{pmatrix} \au \rule[-10pt]{0mm}{0mm} & 0 \\ 
0 & \au^{-1} \end{pmatrix}.
\end{equation*}
Finally, under the action of the gauge group $\underline{U(1)}$ one has
\begin{align*}
U^\au & = U\whS, & \eta^\au &= \eta\,.
\end{align*}

The field $U \in \underline{SU(2)}$ then supports the (commuting) actions of $\underline{SU(2)}$ and $\underline{U(1)}$ respectively by left multiplication and by right multiplication in $SU(2)$. Here $U(1)$ is embedded (through $\whS$) into $SU(2)$ in the $\sigma_3$-direction of the group $SU(2)$.

Using the new variables $U$ and $\eta$ in place of $\phi$, a straightforward computation leads to
\begin{align*}
D_\mu \phi &= (\partial_\mu - i \tfrac{g}{2} b_\mu - i \tfrac{g'}{2} a_\mu) U \eta \spmatrix{0 \\ 1}\\
 &= U (\partial_\mu - i \tfrac{g}{2} B_\mu - i \tfrac{g'}{2} a_\mu) \eta \spmatrix{0 \\ 1}
\end{align*}
where
\begin{equation}
\label{eq-bBrelation}
B_\mu = U^{-1} b_\mu U + \tfrac{2i}{g} U^{-1} (\partial_\mu U)\,.
\end{equation}
Let us stress that the vector field $B_\mu$ \emph{is not obtained by a gauge transformation from the gauge field $b_\mu$} \footnote{This is a known fact that the potential singularities of $U$ at $x$ such that $\eta(x) = 0$ can prevent this relation to be a gauge transformation, which indeed requires a \emph{smooth} function with values in $SU(2)$. Nevertheless, the argument we stress here is at a more fundamental level.}.

Firstly, this is obvious from the fact the $U$ fields are some of the dynamical variables of the theory: a gauge transformation would require a \emph{fixed} element in $\underline{SU(2)}$, but, as dynamical variables, the function $U$ is not such a fixed element. Secondly, a gauge transformation should be applied at the same time on gauge fields and to any scalar and spinor fields coupled to them. In the present situation, the scalar fields $\phi$ are not subject to a transformation accompanying \eqref{eq-bBrelation}. $\phi$ is written in terms of $U$ but it will not be transformed any further. Finally, the following actions of the gauge groups $\underline{U(1)}$ and $\underline{SU(2)}$ on $B_\mu$ prove that $B_\mu$ \emph{is no more a gauge potential for $\underline{SU(2)}$}. This rules out the fact that \eqref{eq-bBrelation} can be a gauge transformation.

Indeed, for the action of $\au \in \underline{U(1)}$, one gets
\begin{align*}
B^\au_\mu &= (U^\au)^{-1} b^\au_\mu U^\au + \tfrac{2i}{g} (U^\au)^{-1} (\partial_\mu (U^\au))\\
&= \whS^{-1} U^{-1} b_\mu U \whS + \tfrac{2i}{g} \whS^{-1} U^{-1} [(\partial_\mu U) \whS + U (\partial_\mu \whS) ]\\
&= \whS^{-1} B_\mu \whS + \tfrac{2i}{g} \whS^{-1} \partial_\mu \whS\,,
\end{align*}
whereas for the action of $\asu \in \underline{SU(2)}$, one gets
\begin{align*}
B^\asu_\mu &= (U^\asu)^{-1} b^\asu_\mu U^\asu + \tfrac{2i}{g} (U^\asu)^{-1} (\partial_\mu (U^\asu))\\
&= U^{-1} \asu [ \asu^{-1} b_\mu \asu + \tfrac{2i}{g} \asu^{-1} \partial_\mu \asu ] \asu^{-1} U \\
&\phantom{===} + \tfrac{2i}{g} U^{-1} \asu [(\partial_\mu \asu^{-1}) U + \asu^{-1} (\partial_\mu U) ]\\
&= B_\mu\,.
\end{align*}

Developing $B_\mu$ as $B_\mu = B_\mu^a \sigma_a$ and defining $W^\pm_\mu = \tfrac{1}{\sqrt{2}}(B^1_\mu \mp i B^2_\mu)$, such that 
\begin{equation*}
B_\mu = \begin{pmatrix} 
B^3_\mu \rule[-10pt]{0mm}{0mm} & \sqrt{2} W^+_\mu \\ 
\sqrt{2} W^-_\mu & -B^3_\mu
\end{pmatrix},
\end{equation*}
one gets
\begin{equation*}
B^\au_\mu = \begin{pmatrix} 
B^3_\mu + \tfrac{2i}{g} \au^{-1} \partial_\mu \au \rule[-10pt]{0mm}{0mm} & \au^{-2} \sqrt{2} W^+_\mu \\ 
\au^{2} \sqrt{2} W^-_\mu & -(B^3_\mu + \tfrac{2i}{g} \au^{-1} \partial_\mu \au)
\end{pmatrix},
\end{equation*}
from which we deduce the transformations of these new fields under the action of the two gauge groups:
\begin{gather}
(W^+_\mu)^\asu = W^+_\mu
\quad\quad\quad
(W^-_\mu)^\asu = W^-_\mu
\quad\quad\quad
(B^3_\mu)^\asu = B^3_\mu
\nonumber\\
\label{eq-gaugeWau}
(W^+_\mu)^\au = \au^{-2} W^+_\mu
\quad\quad\quad
(W^-_\mu)^\au = \au^{2} W^-_\mu
\\
\label{eq-gaugeBau}
(B^3_\mu)^\au = B^3_\mu + \tfrac{2i}{g} \au^{-1} \partial_\mu \au\,.
\end{gather}
The $B_\mu$ fields, and so the $W^\pm_\mu$ and $B^3_\mu$ fields, are \emph{invariant} under the action of the gauge group $\underline{SU(2)}$. The $W^\pm_\mu$ fields are $U(1)$-charged with opposite charges, whereas the $B^3_\mu$ fields behave like a $\underline{U(1)}$-gauge potential.

Let us now look at the fermion fields. Let us define $\psi_L' = U^{-1} \psi_L$ and $\psi_R' = \psi_R$, with $\psi_L'= \spmatrix{\nu_L \\ e_L}$ and $\psi_R' = e_R$. Then one obviously gets
\begin{align*}
D^L_\mu \psi_L &= U (\partial_\mu - i \tfrac{g}{2} B_\mu + i \tfrac{g'}{2} a_\mu) \psi_L'
\\
& \equiv U {D'}^{L}_\mu \psi_L' \\
D^R_\mu \psi_R &= (\partial_\mu + i g' a_\mu) \psi_R' = D^R_\mu \psi_R' \\
f (\overline{\psi_L} \phi \psi_R + \overline{\psi_R} \phi^\dagger \psi_L) &= f \eta (\overline{e_L} e_R + \overline{e_R} e_L)\,.
\end{align*}

A simple computation shows that the new fermions fields have the following transformation rules under the gauge groups actions:
\begin{align}
\label{eq-gaugenueau}
\nu_L^\au &= \nu_L
&
e_L^\au &= \au^2 e_L
&
e_R^\au &= \au^2 e_R\\
\nu_L^\asu &= \nu_L
&
e_L^\asu &= e_L
&
e_R^\asu &= e_R \nonumber
\,.
\end{align}

Finally, with the usual definitions $\cos \theta_W = \tfrac{g}{\sqrt{g^2 + g'^2}}$ and $\sin \theta_W = \tfrac{g'}{\sqrt{g^2 + g'^2}}$, the vector fields
\begin{align*}
Z_\mu &= \cos \theta_W B^3_\mu - \sin \theta_W a_\mu
\\
A_\mu &= \sin \theta_W B^3_\mu + \cos \theta_W a_\mu
\end{align*}
have the transformation rules
\begin{align}
\label{eq-gaugeZAau}
Z_\mu^\au &= Z_\mu
&
A_\mu^\au &= A_\mu + 2 i \frac{1}{e} \au^{-1} \partial_\mu \au\\
Z_\mu^\asu &= Z_\mu
&
A_\mu^\asu &= A_\mu \nonumber
,
\end{align}
where the electric charge $e$ is defined as $e=g \sin \theta_W$.

A standard computation then leads to the following form of the Lagrangian~\eqref{eq-lagrangianbeginning}:
\begin{multline}
\label{eq-lagrangiannewvariables}
\mathcal{L} = (\partial_\mu \eta)(\partial^\mu \eta) - \mu^2 \eta^2 - \lambda \eta^4\\
+ \eta^2(\tfrac{g^2+ g'^2}{4} Z_\mu Z^\mu + \tfrac{g^2}{2}W^+_\mu W^{-\,\mu}) + \eta f (\overline{e_L} e_R+ \overline{e_R} e_L)\\
 + \overline{\psi'_L} i \gamma^\mu {D'}^{L}_\mu \psi_L' + \overline{\psi'_R} i \gamma^\mu D^R_\mu \psi_R'\\
   -\frac{1}{4} f_{\mu \nu} f^{\mu \nu} -\frac{1}{4} \sum_a G^a_{\mu \nu} G^{a\; \mu \nu}
\end{multline}
where $G_{\mu \nu}$  is the ``field strength'' of $B_\mu$, \textit{i.e.} the same expression defining the field strength of an ordinary non abelian gauge field applied to $B_\mu$. Notice that as for a true gauge transformation, using the property of the trace, one has
\begin{equation*}
-\frac{1}{4} \sum_a g^a_{\mu \nu} g^{a\; \mu \nu} = -\frac{1}{4} \sum_a G^a_{\mu \nu} G^{a\; \mu \nu}\,.
\end{equation*}

As usual, the definition of the $Z_\mu$ fields is the only linear combination which compensates the two inhomogeneous terms in the $\underline{U(1)}$-gauge transformations of $B^3_\mu$ and $a_\mu$. Similarly, the definition of $A_\mu$ is submitted to the requirement that the quadratic part in the fields $B^3_\mu$ and $a_\mu$ in the terms $-\frac{1}{4} f_{\mu \nu} f^{\mu \nu} -\frac{1}{4} G^a_{\mu \nu} G_a^{\mu \nu}$ remains diagonal (in the sense that the quadratic parts in the fields $Z_\mu$ and $A_\mu$ are decoupled). This imposes to complete the definition of $Z_\mu$ as a real rotation on the real $2$-dimensional vector $\spmatrix{B^3_\mu\\ a_\mu}$.

The rest of the computation is standard, and gives rise to the (corresponding part of the) Lagrangian of the Standard Model in what is called the broken phase.

It follows that the Lagrangian \eqref{eq-lagrangiannewvariables} is invariant (as a Lagrangian) under $\underline{U(1)}$-gauge transformations, so that it is a $\underline{U(1)}$-gauge field theory. But the most important fact is that it is completely written in term of \emph{invariant fields under $\underline{SU(2)}$-gauge transformations} and it does not depend anymore on the $U$ fields which are the only fields to still support non invariant transformations rules under the action of $\underline{SU(2)}$. This means that the gauge group $\underline{SU(2)}$ has been \emph{decoupled} as a natural symmetry of this model, in the sense that it does not induce any active transformations on any component of the theory.

The mentioned singularity at $\eta=0$ is removed from the Lagrangian \eqref{eq-lagrangiannewvariables} because the only potentially singular fields, $U$, are no more variables of the theory. This means that this Lagrangian makes sense also for $\eta=0$.

\section{Discussion}

Let us comment this computation and show how it is different from the usual one presented in textbooks.

Let us recall the essential steps of the ``standard procedure''. One first writes the scalar fields $\phi$ in the unitary gauge as in~\eqref{eq-phiUeta}. Then, for $\mu^2 < 0$, one finds the minima of the potential $V(\phi) = \mu^2 \phi^\dagger \phi + \lambda (\phi^\dagger \phi)^2$, which are of the form $|\phi| = \frac{v}{\sqrt{2}}$ for $v = \sqrt{-\mu^2/\lambda}$. In the variables $(\eta, U)$, this implies that $\eta = \frac{v}{\sqrt{2}}$ and that $U$ can be \emph{any function} with values in $SU(2)$. This $U$ contains the symmetry of the fundamental configuration of the fields $\phi$. In a quantum approach, this would be the vacuum of the theory.

Symmetry breaking enters then the scene to \emph{fix by hand a particular value to the $U$ fields}. Then, performing a \emph{true} gauge transformation with $u = U$ (in our notations) yields essentially the same computation as the one presented here. We stress the fact that such a gauge transformation is only possible when the $U$ fields are no longer dynamical variables. Indeed, if they were some of the new variables which parametrize the $\phi$ fields, then no gauge transformation with a \emph{fixed} gauge group element $u = U$ would be possible.

The $v$ parameter inserts mass terms in the Lagrangian written in terms of the gauge transformed fields.

In this ``standard procedure'', the obtained Lagrangian does not support a $\underline{SU(2)}$-gauge action because this symmetry is \emph{forbidden} to act on the scalar fields $\phi$, which are required to be written in the chosen form $\phi = \frac{1}{\sqrt{2}}\spmatrix{0 \\ v + \eta'}$. This is indeed the meaning of the fact that $U$ is fixed once and for all in order that it can be removed by a true gauge transformation.

\medskip
In our computation, there is a \emph{clear separation between what happens to fields variables and the genesis of mass terms}. 

Indeed, the requirement $\mu^2 < 0$ is not necessary to obtain and to reveal the field content in terms of photon fields ($A_\mu$), charged and neutral current fields ($W_\mu^\pm$, $Z_\mu$), neutrino and electron fields ($\nu_L$, $e_L$, $e_R$). The Lagrangian obtained in \eqref{eq-lagrangiannewvariables} is written for any values of $\mu^2$. 

As a consequence, the Lagrangian \eqref{eq-lagrangiannewvariables} does contain two phases: one where the fields $W_\mu^\pm$, $Z_\mu$ and $e_L + e_R$ get masses and one where these fields remain massless.

In the phase $\mu^2 < 0$, finding the minimum of the potential $V(\eta) = \mu^2 \eta^2 + \lambda \eta^4$ of the field $\eta$ in \eqref{eq-lagrangiannewvariables} yields the \emph{unique solution} of the form $\eta = \frac{v}{\sqrt{2}}$. It is unique because $\eta$ is constrained to be real and positive. It is then convenient to perform a development of $\eta$ around this value in the form $\eta = \frac{v}{\sqrt{2}} + H$, where $H$ is a new field identified as the scalar Higgs field. This procedure then inserts mass terms at various places in the Lagrangian as expected. 

In the phase $\mu^2 \geq 0$, the Lagrangian does not contain any massive fields except for the $\eta$ field (when $\mu^2 \neq 0$).

Notice that here the genesis of mass terms is not really due to a ``spontaneous symmetry breaking mechanism'' because this minimum, when it exists with a non zero value ($\mu^2 < 0$), is now unique. This is just the usual procedure in fields theory to develop the Lagrangian around the fundamental configurations of the fields, the $\eta$ field in the present case for which there is a unique fundamental configuration.

Notice also that any transition from the phase $\mu^2 < 0$ to the phase $\mu^2 \geq 0$ (and vice-versa) does not change the fields content of the theory except for the scalar Higgs field $H$.

It is worth mentioning that the (three) degrees of freedom of the $U$ fields are exactly the usual (three) Goldstone modes. This computation shows clearly how these Goldstone modes are combined with the $b_\mu$ fields and with the fermions, so that all the new fields involved in \eqref{eq-lagrangiannewvariables} are  $SU(2)$-gauge invariant. In particular, this permits the vector currents to appear in some manifestly non gauge invariant terms like $\eta^2(\tfrac{g^2+ g'^2}{4} Z_\mu Z^\mu + \tfrac{g^2}{2}W^+_\mu W^{-\,\mu})$, which, after choosing a non zero fundamental configuration for $\eta$ if it exists, gives masses to theses vector currents. In that respect, some of the technical considerations presented here bear similarities with some computations performed in the context of non-linear sigma models where Goldstone modes are treated in an analogous way (see \cite{Becc:88a, *Zinn:90} for instance) or more generally in the context of non-linearly realized gauge groups (see \cite{BettFerrQuad:08b, *BettFerrQuad:08c, *BettFerrQuad:08a} and references therein for instance).

\medskip
In our computation, the relation $J^{\text{em}} = J^3 + \frac{1}{2} J^{Y}$ between the electromagnetic current $J^{\text{em}}$, the $SU(2)$ current $J^3$ and the hypercharge current $J^{Y}$ is hidden at various places. The model has indeed been built on this relation. The interpretation of the fields content of the Lagrangian~\eqref{eq-lagrangiannewvariables} in terms of particles relies heavily on the quantum numbers one can associate to these fields. For instance, the electromagnetic charge operator, written usually as $Q = T^3 + \frac{1}{2} Y$ with obvious notations, identifies the components of $\psi'_L$ as a neutral and a charged particle. An other way to read off these quantum numbers is to look at gauge transformations. Equations~\eqref{eq-gaugeWau}, \eqref{eq-gaugeBau} , \eqref{eq-gaugenueau} and \eqref{eq-gaugeZAau} show that $A_\mu$ is the $\underline{U(1)}$-gauge potential associated to the following charge particles: $W^+_\mu$ is a vector fields of charge $-e$, $W^-_\mu$ is a vector fields of charge $+e$, $e_L$ and $e_R$ have charge $+e$ and $Z_\mu$ is a neutral vector fields as is $\nu_L$. This is the electromagnetism $\underline{U(1)}$-gauge theory in the ``broken'' phase. 

We stress the fact that contrary to the ordinary procedure, the final gauge group $\underline{U(1)}$ which defines the remaining electromagnetism gauge theory is the $\underline{U(1)}$ part of the gauge group $\underline{U(1)} \times \underline{SU(2)}$.

The relation $J^{\text{em}} = J^3 + \frac{1}{2} J^{Y}$ is a consequence of the choice of the unit real vector $\spmatrix{0 \\ 1}$ in equation~\eqref{eq-phiUeta}, and can be traced in the computation up to the assigned $U(1)$ charges of the new fields. In particular, the matrix $\whS$ is an essential step in this process, when one looks at it as an embedding of $U(1)$ into $SU(2)$.

\medskip
\emph{The choice of the unit real vector $\spmatrix{0 \\ 1}$ is related to the requirement that the final $U(1)$ charges must be clearly identified}. Indeed, we can modify equation~\eqref{eq-phiUeta} in its most general expression
\begin{equation*}
\label{eq-phiUetaalpha}
\phi = U(v) \eta\, v \begin{pmatrix} 0 \\ 1 \end{pmatrix},
\end{equation*}
where $v \in SU(2)$ is a constant matrix which defines a \emph{constant} reference vector $v \spmatrix{0 \\ 1} \in \gC^2$ of norm $1$. Here $U(v) \in \underline{SU(2)}$ depends on this matrix $v$ but $\eta$ (as the norm of $\phi$) does not, so that it is the same as in \eqref{eq-phiUeta}. For $v = \bbbone_2$, one recovers the situation discussed before. A direct comparison leads to $U(v) = U v^{-1}$, which implies, with previous notations,
\begin{align*}
U(v)^\asu &= \asu^{-1} U(v),
&
U(v)^\au &= U(v) \whS(v),
\end{align*}
with $\whS(v) = v \whS v^{-1}$.

In order to factor out $U(v)$ in $D_\mu \phi$, one introduces the new fields
\begin{equation*}
B(v)_\mu = U(v)^{-1} b_\mu U(v) + \tfrac{2i}{g} U(v)^{-1} (\partial_\mu U(v))\,.
\end{equation*}
It is easy to see that $B(v)_\mu = v B_\mu v^{-1}$, so that
\begin{align*}
B(v)^\asu_\mu &= B(v)_\mu
&
B(v)^\au_\mu &= v B^\au_\mu v^{-1}\,. 
\end{align*}
Notice that these relations use the fact that $v$ is a constant matrix.

The new fermion fields are defined in the same way as $\psi(v)_L = U(v)^{-1} \psi_L$ and $\psi(v)_R = \psi_R$. One then has
\begin{equation*}
\psi(v)^\au_L = v
\begin{pmatrix}
1 & 0 \\ 
0 & \au^2
\end{pmatrix}
v^{-1} \psi(v)_L\,.
\end{equation*}
In order that the two components of the field $\psi(v)_L$ behave as clearly identified $U(1)$-charged particles, it is necessary to have $v \spmatrix{1 & 0 \\ 0 & \au^2} v^{-1}$ diagonal. This is achieved only when
\begin{equation*}
v = 
\begin{pmatrix}
e^{i \alpha} & 0 \\
0 & e^{-i \alpha}
\end{pmatrix}
\ \ \ \text{ or }\ \ \ 
v = 
\begin{pmatrix}
0 & -e^{-i \beta} \\
e^{i \beta} & 0 
\end{pmatrix} 
\end{equation*}
for constants $\alpha, \beta \in \gR$. The case $v = \bbbone_2$ corresponds to $\alpha \equiv 0 \bmod 2 \pi$.

For any constant values $\alpha, \beta \in \gR$, the field content of the obtained Lagrangians is exactly the same. These different theories are related by global transformations involving constant phases. For instance, using previous notations, one has $\psi(v)_L = v \psi_L' = v \spmatrix{\nu_L \\ e_L}$, which relates fermions fields, and the previously mentioned relation $B(v)_\mu = v B_\mu v^{-1}$, which gets rid of the fields $Z_\mu$ and $W_\mu^\pm$. Notice that if one does not require that the final fields behave correctly under $\underline{U(1)}$-gauge transformations, any values of $v$ are acceptable.

\medskip
The computation we have presented here is only based on a convenient change of variables at the classical level of the theory, given essentially by \eqref{eq-phiUeta} and \eqref{eq-bBrelation}. It does not make any reference to some ``broken vacuum'' (fundamental configuration of $\phi$) which would be responsible for the (spontaneous) breaking of the symmetry.

The most important outcome is that at the classical level the $\underline{SU(2)}$ symmetry is not broken: it is in fact factored out from the Lagrangian by this change of variables, as one would have expected in some gauge fixing procedure. The gauge group $\underline{SU(2)}$ is always there, but it \emph{does not effectively operate anymore on any fields}. This really changes the point of view one has to have on this mechanism in the Standard Model: it looks more like an efficient gauge fixing procedure performed at the classical level than a symmetry breaking mechanism.

It is well known that the quantization of the Standard Model is well behaved with the full Lagrangian~\eqref{eq-lagrangianbeginning}, for which renormalizability has been established with success, using for instance the $R_\xi$-gauge \cite{t-Ho:71a, *t-Ho:71b, *t-HoVelt:72a, *t-HoVelt:72b}. On the contrary, the unitary gauge which is used in our computation is not suitable to construct a reliable perturbative renormalizable theory. Our computation does not seem to give any new light on the technical aspect of the quantization of the theory.

Indeed, one can perform the change of variables presented here in the functional integral defining the quantized version of the theory, starting from the Lagrangian~\eqref{eq-lagrangianbeginning}. Using the Jacobian of the change of variables from $(\phi, b_\mu, \psi_L)$ to $(U, \eta, B_\mu, \psi_L')$ (see Appendix~\ref{sec-computationjacobian}), one has
\begin{multline*}
\int [d\phi] [d b_\mu] [d a_\mu] [d \psi_L] [d \psi_R] e^{iS[\phi, b_\mu, a_\mu, \psi_L, \psi_R]}
= \\
\int \eta^3 [d\eta] [d U] [d B_\mu] [d a_\mu] [d \psi_L'] [d \psi_R'] e^{iS[\eta, B_\mu, a_\mu, \psi_L', \psi_R']}
\end{multline*}

On the one hand, in the second expression the integration along the $U$ fields over the gauge group $\underline{SU(2)}$ can be factored out because the action does not depend anymore of these fields. One step further, this remains true for the functional integral in the variables of the final Lagrangian~\eqref{eq-lagrangiannewvariables}, where the fields $W_\mu^\pm$, $Z_\mu$ and $A_\mu$ have been introduced. The ``volume'' of the gauge group $\underline{SU(2)}$ can then be factored out in the functional integral at the quantum level, as required by a gauge fixing procedure.

But on the other hand, the presence of the term $\eta^3$ in the functional integral measure may be the sign that the quantization of the Higgs sector is non trivial with this final Lagrangian.

\appendix

\section{Computation of the Jacobian}
\label{sec-computationjacobian}

For sake of completeness, we expose here the computation of the Jacobian arising in the functional integral for the change of variables $(\phi, b, \psi_L) \mapsto (U, \eta, B, \psi_L')$, where $b = (b_\mu)$ and $B = (B_\mu)$. In fact, it is convenient to compute this Jacobian for the new variables $(U, \sigma, B, \psi_L')$ where the new field $\sigma$ is defined as $\eta = e^\sigma$. At the end, we will replace $\eta [d\sigma]$ by $[d\eta]$.

The change of variables between these fields is summarized in the following relations, where old fields are expressed in terms of new fields:
\begin{gather}
\label{eq-newvariablesoldvariables1}
\phi = U e^\sigma \begin{pmatrix} 0 \\ 1 \end{pmatrix},
\ \ \ \ \
\psi_L = U \psi_L',
\\
\label{eq-newvariablesoldvariables2}
b_\mu = U B_\mu U^{-1} + \tfrac{2i}{g} U (\partial_\mu U^{-1})
\end{gather}

The functional spaces in which these fields are defined are given by:
\begin{align*}
\phi &\in \underline{\gC^2}
&
b, B &\in \caA
&
\psi_L, \psi_L' &\in \caS \otimes \gC^2
\\
U & \in \underline{SU(2)}
&
\sigma &\in \underline{\gR}
&
 &
\end{align*}
where underlined symbols are spaces of functions with values in the specified space, $\caA$ is the space of $\ksu(2)$-connection $1$-forms and $\caS$ is the space of spinors.

A small variation of $U \in \underline{SU(2)}$ is an element $U(\varepsilon) \in \underline{SU(2)}$, where $\varepsilon$ is a small real parameter, such that $U(0) = U$. The quantity $U(\varepsilon) U^{-1}$ is closed to the identity in $\underline{SU(2)}$, and can be parametrized as $U(\varepsilon) U^{-1} = e^{-i \varepsilon \alpha}$ for a function $\alpha \in \underline{\ksu(2)}$. Deriving along $\varepsilon$, one gets, with the notation $\delta U = \frac{d U(\varepsilon)}{d \varepsilon}_{| \varepsilon = 0}$, $(\delta U) U^{-1} = - i \alpha$. This gives also $U (\delta U^{-1}) = i \alpha$.

A small variation of $\sigma$ is a $\gR$-valued function $\delta \sigma = \lambda \in \underline{\gR}$. A small variation of $\phi$ is an element $\delta \phi \in \underline{\gC^2}$.

A small variation of the fields $B$ is a collection of functions $\beta_\mu \in \underline{\ksu(2)}$. In a more geometric language, $\beta$ defines an element in $\Omega^1 \otimes \ksu(2)$, the space of $1$-forms on space-time with values in the Lie algebra $\ksu(2)$. This is the tangent space to the space $\caA$ of $\ksu(2)$-connection $1$-forms. We introduce the notation $\delta B = \beta$. A small variation of $b$ takes place in the same space.

Small variations of $\psi_L$ and $\psi_L'$ are functions in $\caS \otimes \gC^2$. We will denote by $\delta \psi_L' = \varphi$ such a variation.

The Jacobian one has to compute is the determinant of the matrix
\begin{equation*}
J(U, \eta, B, \psi_L') = 
\begin{pmatrix}
\frac{\delta \phi}{\delta U} & \frac{\delta \phi}{\delta \sigma} & \frac{\delta \phi}{\delta B} & \frac{\delta \phi}{\delta \psi_L'}
\\
\frac{\delta b}{\delta U} & \frac{\delta b}{\delta \sigma} & \frac{\delta b}{\delta B} & \frac{\delta b}{\delta \psi_L'}
\\
\frac{\delta \psi_L}{\delta U} & \frac{\delta \psi_L}{\delta \sigma} & \frac{\delta \psi_L}{\delta B} & \frac{\delta \psi_L}{\delta \psi_L'}
\end{pmatrix}
\end{equation*}

Using equations \eqref{eq-newvariablesoldvariables1} and \eqref{eq-newvariablesoldvariables2}, straightforward computations give
\begin{align*}
\delta \phi &= - (U \delta U^{-1}) U e^\sigma \begin{pmatrix} 0 \\ 1 \end{pmatrix} + U e^\sigma (\delta \sigma)\begin{pmatrix} 0 \\ 1 \end{pmatrix}
\\
\delta b &= \frac{2i}{g} \left( d (U \delta U^{-1}) + \tfrac{g}{2i}\left[ b , U \delta U^{-1} \right]  \right) + U (\delta B) U^{-1}
\\
\delta \psi_L &= - (U \delta U^{-1}) U \psi_L' + U \delta \psi_L'
\end{align*}
This implies that the only non zero blocks in $J$ are given by the following maps:
\begin{align*}
\tfrac{\delta \phi}{\delta U} &: \underline{\ksu(2)}  \rightarrow \underline{\gC^2}
&
\alpha &\mapsto -i \alpha \phi
\\
\tfrac{\delta \phi}{\delta \sigma} &: \underline{\gR}  \rightarrow \underline{\gC^2}
&
\lambda &\mapsto \lambda \phi
\\
\tfrac{\delta b}{\delta U} &: \underline{\ksu(2)}  \rightarrow \Omega^1 \otimes \ksu(2)
&
\alpha &\mapsto \tfrac{2i}{g} D_b \alpha
\\
\tfrac{\delta b}{\delta B} &: \Omega^1 \otimes \ksu(2) \rightarrow \Omega^1 \otimes \ksu(2)
&
\beta &\mapsto \Ad_U \beta
\\
\tfrac{\delta \psi_L}{\delta U} &: \underline{\ksu(2)}  \rightarrow \caS \otimes \gC^2
&
\alpha &\mapsto -i \alpha \psi_L
\\
\tfrac{\delta \psi_L}{\delta \psi_L'} &: \caS \otimes \gC^2 \rightarrow \caS \otimes \gC^2
&
\varphi &\mapsto U \varphi
\end{align*}
where $D_b$ is the covariant derivative for the connection $b$ and $\Ad_U \beta = U \beta U^{-1}$.

The block $\left( \tfrac{\delta \phi}{\delta U} \tfrac{\delta \phi}{\delta \sigma}  \right)$ is a square matrix in the sense that it is a map $\underline{\gR}^4  \rightarrow \underline{\gR}^4$ when we identify $\underline{\gR}^3 \simeq \underline{\ksu(2)}$ and $\underline{\gR}^4 \simeq \underline{\gC^2}$. The maps $\tfrac{\delta b}{\delta B}$ and $\tfrac{\delta \psi_L}{\delta \psi_L'}$ are also ``square matrices'' in the same way. The matrix $J$ is then lower triangular, so that its determinant is just the product of the determinant of these three blocks.

The map $\left( \tfrac{\delta \phi}{\delta U} \tfrac{\delta \phi}{\delta \sigma}  \right)$ is linear on functions, the map $\tfrac{\delta b}{\delta B}$ is linear on $1$-forms and the map $\tfrac{\delta \psi_L}{\delta \psi_L'}$ is linear on spinors. It is then easy to see that their determinants can be computed as determinants in some finite dimensional vector spaces at each point in space-time: $\gR^4$ for $\left( \tfrac{\delta \phi}{\delta U} \tfrac{\delta \phi}{\delta \sigma}  \right)$, $\ksu(2)$ for $\tfrac{\delta b}{\delta B}$ and $\gC^2$ for $\tfrac{\delta \psi_L}{\delta \psi_L'}$.

The determinant of $\varphi \mapsto U \varphi$ on $\gC^2$ is obviously $1$ for any $U$. In the same way, results in linear algebra and Lie algebras show that the determinant of $\beta \mapsto \Ad_U \beta$ is $1$ on $\ksu(2)$. In the basis $\sigma_1, \sigma_2, \sigma_3, \bbbone$ of $\ksu(2) \oplus \gR \simeq \gR^4$ and the basis $\spmatrix{1 \\ 0}, \spmatrix{i \\ 0}, \spmatrix{0 \\ 1}, \spmatrix{0 \\ i}$ of $\gC^2 \simeq \gR^4$, the map $\left( \tfrac{\delta \phi}{\delta U} \tfrac{\delta \phi}{\delta \sigma}  \right)$ takes the explicit form of the $4 \times 4$ matrix
\begin{equation*}
\begin{pmatrix}
 \phi_{2y} & -\phi_{2x} &  \phi_{1y} &  \phi_{1x} \\
-\phi_{2x} & -\phi_{2y} & -\phi_{1x} &  \phi_{1y} \\
 \phi_{1y} &  \phi_{1x} & -\phi_{2y} &  \phi_{2x} \\
-\phi_{1x} &  \phi_{1y} &  \phi_{2x} &  \phi_{2y} 
\end{pmatrix},
\end{equation*}
where $\phi = \spmatrix{\phi_{1x} + i \phi_{1y} \\ \phi_{2x} + i \phi_{2y}}$. The determinant of this matrix is $(\phi_{1x}^2 + \phi_{1y}^2 + \phi_{2x}^2 + \phi_{2y}^2)^2 = \eta^4$.

The determinant of $J$ is then $\eta^4$, which shows that the functional integral measures are related by
\begin{align*}
[d\phi] [d b] [d \psi_L] &= \eta^4 [d \sigma] [d U] [d B] [d \psi_L'] 
\\
&= \eta^3 [d \eta] [d U] [d B] [d \psi_L']
\end{align*}

\begin{acknowledgments}
We would like to thank A.~Abada, S.~Lazzarini and T.~Schücker for stimulating discussions.
\end{acknowledgments}

\bibliography{biblio-articles-perso,biblio-livre,biblio-articles}

\begin{thebibliography}{10}%
\makeatletter
\providecommand \@ifxundefined [1]{%
 \ifx #1\undefined \expandafter \@firstoftwo
 \else \expandafter \@secondoftwo
\fi
}%
\providecommand \@ifnum [1]{%
 \ifnum #1\expandafter \@firstoftwo
 \else \expandafter \@secondoftwo
\fi
}%
\providecommand \enquote [1]{``#1''}%
\providecommand \bibnamefont  [1]{#1}%
\providecommand \bibfnamefont [1]{#1}%
\providecommand \citenamefont [1]{#1}%
\providecommand\href[0]{\@sanitize\@href}%
\providecommand\@href[1]{\endgroup\@@startlink{#1}\endgroup\@@href}%
\providecommand\@@href[1]{#1\@@endlink}%
\providecommand \@sanitize [0]{\begingroup\catcode`\&12\catcode`\#12\relax}%
\@ifxundefined \pdfoutput {\@firstoftwo}{%
 \@ifnum{\z@=\pdfoutput}{\@firstoftwo}{\@secondoftwo}%
}{%
 \providecommand\@@startlink[1]{\leavevmode\special{html:<a href="#1">}}%
 \providecommand\@@endlink[0]{\special{html:</a>}}%
}{%
 \providecommand\@@startlink[1]{%
  \leavevmode
  \pdfstartlink
   attr{/Border[0 0 1 ]/H/I/C[0 1 1]}%
   user{/Subtype/Link/A<</Type/Action/S/URI/URI(#1)>>}%
  \relax
 }%
 \providecommand\@@endlink[0]{\pdfendlink}%
}%
\providecommand \url  [0]{\begingroup\@sanitize \@url }%
\providecommand \@url [1]{\endgroup\@href {#1}{\urlprefix}}%
\providecommand \urlprefix [0]{URL }%
\providecommand \Eprint[0]{\href }%
\@ifxundefined \urlstyle {%
  \providecommand \doi [1]{doi:\discretionary{}{}{}#1}%
}{%
  \providecommand \doi [0]{doi:\discretionary{}{}{}\begingroup
  \urlstyle{rm}\Url }%
}%
\providecommand \doibase [0]{http://dx.doi.org/}%
\providecommand \Doi[1]{\href{\doibase#1}}%
\providecommand \bibAnnote [3]{%
  \BibitemShut{#1}%
  \begin{quotation}\noindent
    \textsc{Key:}\ #2\\\textsc{Annotation:}\ #3%
  \end{quotation}%
}%
\providecommand \bibAnnoteFile [2]{%
  \IfFileExists{#2}{\bibAnnote {#1} {#2} {\input{#2}}}{}%
}%
\providecommand \typeout [0]{\immediate \write \m@ne }%
\providecommand \selectlanguage [0]{\@gobble}%
\providecommand \bibinfo [0]{\@secondoftwo}%
\providecommand \bibfield [0]{\@secondoftwo}%
\providecommand \translation [1]{[#1]}%
\providecommand \BibitemOpen[0]{}%
\providecommand \bibitemStop [0]{}%
\providecommand \bibitemNoStop [0]{.\EOS\space}%
\providecommand \EOS [0]{\spacefactor3000\relax}%
\providecommand \BibitemShut [1]{\csname bibitem#1\endcsname}%
\bibitem{Glas:61a}%
  \BibitemOpen
  \bibfield{author}{%
  \bibinfo {author} {\bibfnamefont{S.~L.}\ \bibnamefont{Glashow}},\ }%
  \bibfield{journal}{%
  \bibinfo {journal} {Nucl. Phys.}\ }%
  \textbf{\bibinfo {volume} {22}},\ \bibinfo {pages} {579} (\bibinfo {year}
  {1961})%
  \bibAnnoteFile{NoStop}{Glas:61a}%
\bibitem{Wein:67a}%
  \BibitemOpen
  \bibfield{author}{%
  \bibinfo {author} {\bibfnamefont{S.}~\bibnamefont{Weinberg}},\ }%
  \bibfield{journal}{%
  \Doi{10.1103/PhysRevLett.19.1264}{\bibinfo {journal} {Phys. Rev. Lett.}}\ }%
  \textbf{\bibinfo {volume} {19}},\ \bibinfo {pages} {1264} (\bibinfo {month}
  {Nov}\ \bibinfo {year} {1967})%
  \bibAnnoteFile{NoStop}{Wein:67a}%
\bibitem{Sala:68a}%
  \BibitemOpen
  \bibfield{author}{%
  \bibinfo {author} {\bibfnamefont{A.}~\bibnamefont{Salam}},\ }%
  in\ \emph{\bibinfo {booktitle} {Elementary particle theory: relativistic
  groups and analyticity, Proceedings of the 8th {N}obel Symposium}},\ \bibinfo
  {editor} {edited by\ \bibinfo {editor}
  {\bibfnamefont{N.}~\bibnamefont{Svartholm}}}\ (\bibinfo {publisher} {Almquist
  and Forlag},\ \bibinfo {address} {Stockholm},\ \bibinfo {year} {1968})%
  \bibAnnoteFile{NoStop}{Sala:68a}%
\bibitem{Wein:05b}%
  \BibitemOpen
  \bibfield{author}{%
  \bibinfo {author} {\bibfnamefont{S.}~\bibnamefont{Weinberg}},\ }%
  \emph{\bibinfo {title} {The Quantum Theory Of Fields: Modern Applications}}\
  (\bibinfo {publisher} {Cambridge University Press},\ \bibinfo {year} {2005})%
  \bibAnnoteFile{NoStop}{Wein:05b}%
\bibitem{PeskSchr:08a}%
  \BibitemOpen
  \bibfield{author}{%
  \bibinfo {author} {\bibfnamefont{M.~E.}\ \bibnamefont{Peskin}}\ and\ \bibinfo
  {author} {\bibfnamefont{D.~V.}\ \bibnamefont{Schroeder}},\ }%
  \emph{\bibinfo {title} {An Introduction to Quantum Field Theory}}\ (\bibinfo
  {publisher} {Perseus Books},\ \bibinfo {year} {2008})%
  \bibAnnoteFile{NoStop}{PeskSchr:08a}%
\bibitem{AitcHey:04a}%
  \BibitemOpen
  \bibfield{author}{%
  \bibinfo {author} {\bibfnamefont{I.~J.~R.}\ \bibnamefont{Aitchison}}\ and\
  \bibinfo {author} {\bibfnamefont{A.~J.~G.}\ \bibnamefont{Hey}},\ }%
  \emph{\bibinfo {title} {Gauge Theories in Particle Physics - Vol. 2:
  Non-Abelien Gauge Theories: {QCD} and the Electroweak Theory}}\ (\bibinfo
  {publisher} {Institute of Physics Publishing},\ \bibinfo {year} {2004})%
  \bibAnnoteFile{NoStop}{AitcHey:04a}%
\bibitem{ChenLi:84a}%
  \BibitemOpen
  \bibfield{author}{%
  \bibinfo {author} {\bibfnamefont{T.-P.}\ \bibnamefont{Cheng}}\ and\ \bibinfo
  {author} {\bibfnamefont{L.-F.}\ \bibnamefont{Li}},\ }%
  \emph{\bibinfo {title} {Gauge theory of elementary particle physics}}\
  (\bibinfo {publisher} {Oxford University Press},\ \bibinfo {year} {1984})%
  \bibAnnoteFile{NoStop}{ChenLi:84a}%
\bibitem{Note1}%
  \BibitemOpen
  \bibinfo {note} {This is a known fact that the potential singularities of $U$
  at $x$ such that $\eta (x) = 0$ can prevent this relation to be a gauge
  transformation, which indeed requires a \protect \emph {smooth} function with
  values in $SU(2)$. Nevertheless, the argument we stress here is at a more
  fundamental level.}%
  \bibAnnoteFile{Stop}{Note1}%
\bibitem{Becc:88a}%
  \BibitemOpen
  \bibfield{author}{%
  \bibinfo {author} {\bibfnamefont{C.}~\bibnamefont{Becchi}},\ }%
  in\ \emph{\bibinfo {booktitle} {Renormalization of Quantum Field Theories
  with Non-linear Field Transformations}},\ \bibinfo {series} {Lecture Notes in
  Physics}, Vol.\ \bibinfo {volume} {303}\ (\bibinfo {publisher}
  {Springer-Verlag},\ \bibinfo {year} {1988})\ pp.\ \bibinfo {pages} {95--108}%
  \bibAnnoteFile{NoStop}{Becc:88a}%
\bibitem{Zinn:90}%
  \BibitemOpen
  \bibfield{author}{%
  \bibinfo {author} {\bibfnamefont{J.}~\bibnamefont{Zinn-Justin}},\ }%
  \emph{\bibinfo {title} {Quantum Field Theory and Critical Phenomena}},\
  \bibinfo {series} {International Series of Monographs on Physics},
  Vol.~\bibinfo {volume} {77}\ (\bibinfo {publisher} {Oxford Science
  Publications},\ \bibinfo {year} {1990})%
  \bibAnnoteFile{NoStop}{Zinn:90}%
\bibitem{BettFerrQuad:08b}%
  \BibitemOpen
  \bibfield{author}{%
  \bibinfo {author} {\bibfnamefont{D.}~\bibnamefont{Bettinelli}}, \bibinfo
  {author} {\bibfnamefont{R.}~\bibnamefont{Ferrari}},\ and\ \bibinfo {author}
  {\bibfnamefont{A.}~\bibnamefont{Quadri}},\ }%
  \bibfield{journal}{%
  \bibinfo {journal} {Journal of Generalized Lie Theory and Applications}\ }%
  \textbf{\bibinfo {volume} {2}},\ \bibinfo {pages} {122} (\bibinfo {year}
  {2008})%
  \bibAnnoteFile{NoStop}{BettFerrQuad:08b}%
\bibitem{BettFerrQuad:08c}%
  \BibitemOpen
  \bibfield{author}{%
  \bibinfo {author} {\bibfnamefont{D.}~\bibnamefont{Bettinelli}}, \bibinfo
  {author} {\bibfnamefont{R.}~\bibnamefont{Ferrari}},\ and\ \bibinfo {author}
  {\bibfnamefont{A.}~\bibnamefont{Quadri}},\ }%
  \bibfield{journal}{%
  \bibinfo {journal} {Physical Review D}\ }%
  \textbf{\bibinfo {volume} {77}},\ \bibinfo {pages} {045021} (\bibinfo {year}
  {2008})%
  \bibAnnoteFile{NoStop}{BettFerrQuad:08c}%
\bibitem{BettFerrQuad:08a}%
  \BibitemOpen
  \bibfield{author}{%
  \bibinfo {author} {\bibfnamefont{D.}~\bibnamefont{Bettinelli}}, \bibinfo
  {author} {\bibfnamefont{R.}~\bibnamefont{Ferrari}},\ and\ \bibinfo {author}
  {\bibfnamefont{A.}~\bibnamefont{Quadri}},\ }%
  \enquote{\bibinfo {title} {The {$SU(2) \otimes U(1)$} electroweak model based
  on the nonlinearly realized gauge group},}\ \bibinfo {note} {Preprint},\
  \url{http://arxiv.org/abs/0807.3882v3}%
  \bibAnnoteFile{NoStop}{BettFerrQuad:08a}%
\bibitem{t-Ho:71a}%
  \BibitemOpen
  \bibfield{author}{%
  \bibinfo {author} {\bibfnamefont{G.}~\bibnamefont{'t~{H}ooft}},\ }%
  \bibfield{journal}{%
  \bibinfo {journal} {Nucl. Phys. B}\ }%
  \textbf{\bibinfo {volume} {33}},\ \bibinfo {pages} {173} (\bibinfo {year}
  {1971})%
  \bibAnnoteFile{NoStop}{t-Ho:71a}%
\bibitem{t-Ho:71b}%
  \BibitemOpen
  \bibfield{author}{%
  \bibinfo {author} {\bibfnamefont{G.}~\bibnamefont{'t~{H}ooft}},\ }%
  \bibfield{journal}{%
  \bibinfo {journal} {Nucl. Phys. B}\ }%
  \textbf{\bibinfo {volume} {35}},\ \bibinfo {pages} {167} (\bibinfo {year}
  {1971})%
  \bibAnnoteFile{NoStop}{t-Ho:71b}%
\bibitem{t-HoVelt:72a}%
  \BibitemOpen
  \bibfield{author}{%
  \bibinfo {author} {\bibfnamefont{G.}~\bibnamefont{'t~{H}ooft}}\ and\ \bibinfo
  {author} {\bibfnamefont{M.}~\bibnamefont{Veltman}},\ }%
  \bibfield{journal}{%
  \bibinfo {journal} {Nucl. Phys. B}\ }%
  \textbf{\bibinfo {volume} {44}},\ \bibinfo {pages} {189} (\bibinfo {year}
  {1972})%
  \bibAnnoteFile{NoStop}{t-HoVelt:72a}%
\bibitem{t-HoVelt:72b}%
  \BibitemOpen
  \bibfield{author}{%
  \bibinfo {author} {\bibfnamefont{G.}~\bibnamefont{'t~{H}ooft}}\ and\ \bibinfo
  {author} {\bibfnamefont{M.}~\bibnamefont{Veltman}},\ }%
  \bibfield{journal}{%
  \bibinfo {journal} {Nucl. Phys. B}\ }%
  \textbf{\bibinfo {volume} {50}},\ \bibinfo {pages} {318} (\bibinfo {year}
  {1972})%
  \bibAnnoteFile{NoStop}{t-HoVelt:72b}%
\end{thebibliography}%

\end{document}